\documentclass[twocolumn,astrosymb,tighten,twocolappendix]{aastex631}
\usepackage{amsmath}
\usepackage{ulem}

\newcommand{\modelL}  {$\mathcal{L}$} 
\newcommand{\modelH}  {$\mathcal{H}$}

\newcommand{\crir}  {\zeta^{\rm ion}_{{\rm H}_2}}
\newcommand{\HH}    {H$_2$}
\newcommand{\HHHp}  {H$_3^+$}
\newcommand{\ohhdp} {\mbox{o-H$_2$D$^+$}}
\newcommand{\nHH}   {$n({\rm H_2})$}

\newcommand{\CeigO} {C$^{18}$O}
\newcommand{\HtrCOp}{H$^{13}$CO$^+$}
\newcommand{\DCOp}  {DCO$^+$}

\newcommand{\fD}    {$f_{\rm D}$}

\newcommand{\Tdust} {T_{\rm dust}}

\received{March 6, 2023}
\revised{March 25, 2023}
\accepted{\today}

\shorttitle{ALMA CRIR maps}
\shortauthors{Sabatini G. et al. (2023)}
\graphicspath{{./}{figures/}}

\begin{document}

\title{First ALMA maps of cosmic ray ionisation rate in high-mass star-forming regions}

\correspondingauthor{Giovanni Sabatini}
\email{giovanni.sabatini@inaf.it}

\author[0000-0002-6428-9806]{Giovanni~Sabatini}
\affiliation{INAF - Istituto di Radioastronomia - Italian node of the ALMA Regional Centre (It-ARC), Via Gobetti 101, I-40129 Bologna, Italy}

\author[0000-0003-2814-6688]{Stefano~Bovino}
\affiliation{INAF - Istituto di Radioastronomia - Italian node of the ALMA Regional Centre (It-ARC), Via Gobetti 101, I-40129 Bologna, Italy}
\affiliation{Departamento de Astronom\'ia, Facultad Ciencias F\'isicas y Matem\'aticas, Universidad de Concepci\'on, Av. Esteban Iturra s/n Barrio Universitario, Casilla 160, Concepci\'on, Chile}

\author[0000-0002-0528-8125]{Elena~Redaelli}
\affiliation{Max-Planck-Institut f{\"u}r extraterrestrische Physik, Gie{\ss}enbachstra{\ss}e 1, D-85749 Garching bei
M{\"u}nchen, Germany}


\begin{abstract}
Low-energy cosmic rays ($<1$~TeV) are a pivotal source of ionisation of the interstellar \mbox{medium, where} they play a central role in determining the gas chemical composition and drastically influence the formation of stars and planets. Over the last few decades, \HHHp~absorption lines \mbox{observations in diffuse clouds} have provided reliable estimates of the cosmic ray ionisation rate relative to \HH~($\crir$). However, in denser clouds, where stars and planets form, this method is often inefficient due to the lack of \HHHp~rotational transitions. The $\crir$ estimates are, therefore, still provisional in this context and represent one of the least understood components when it comes to defining general models of star and planet formation. In this Letter, we present the first high-resolution maps of the $\crir$ in two high-mass clumps obtained with a new analytical approach recently proposed to estimate the $\crir$ in the \mbox{densest regions of} molecular clouds. We obtain $\langle\crir\rangle$ that span from $3\times10^{-17}$ to $10^{-16}\rm~s^{-1}$, depending on the different distribution of the main ion carriers, in excellent agreement with the most recent cosmic rays propagation models. The cores belonging to the same parental clump show comparable $\crir$, suggesting that the ionisation properties of prestellar regions are determined by global rather than local effects. These results provide important information for the chemical and physical modelling of star-forming regions.
\end{abstract}

\section{Introduction}\label{sec1:intro}
\begin{deluxetable}{lcc}
\savetablenum{1}
\tablewidth{\columnwidth}
\caption{Physical and observed properties of the sample}\label{tab:summary}
\tablehead{
 \multicolumn{3}{c}{(a) Physical Properties~~}\\ 
 & AG351 &  AG354 }
\startdata
(1) RA (ICRS)               & 17$\rm^h$:20$\rm^m$:51$\rm^s$.03   & 17$\rm^h$:35$\rm^m$:12$\rm^s$.03  \\ 
(2) Dec (ICRS)              & -35$\rm^d$:35$\rm^m$:23$\rm^s$.29  & -33$\rm^d$:30$\rm^m$:28$\rm^s$.97 \\
(3) $d_{\odot}$             &      1.3~kpc  &      1.9~kpc \\
(4) $R_{\rm GC}$            &      7.0~kpc  &      7.4~kpc \\ 
(5) $T_{\rm dust}$          &     17~K    &     19~K   \\ 
(6) $M_{\rm gas}$           & 170~M$_\odot$ & 150~M$_\odot$\\   
\tableline
\multicolumn{3}{c}{(b) Achieved Sensitivity ($\langle rms\rangle$)}\\
Tracer & AG351 & AG354\\
\tableline
(1.3~mm)$_{\rm cont.}$ & 0.1~mJy~beam$^{-1}$ & 0.1~mJy~beam$^{-1}$ \\
 \ohhdp                &     300~mK          & 300~mK \\
 \CeigO                &     210~mK          & 220~mK \\
 \DCOp                 &     160~mK          & 160~mK \\
 \HtrCOp               &     130~mK          & 130~mK \\
 $^{12}$C$^{16}$O      &     80~mK\tablenotemark{\small{i}}         & 90~mK\tablenotemark{\small{i}}\\
\tableline
\multicolumn{3}{c}{(c) Molecular Outflows}\\
\tableline
$^{12}$C$^{16}$O & no & yes
\enddata
\tablecomments{$^{(i)}$ obtained from spectra at \mbox{$\Delta\nu = 0.6$~km~s$^{-1}$}; Upper-panel (a): (1-2) ICRS phase center for ALMA pointings. (3) heliocentric distance; (4) galactocentric distance; (5) dust temperature; (6) total clump's mass. Central panel (b): averaged root mean square (rms) noise for each tracer. Lower panel (c): possible existence of molecular outflows (see Appendix~\ref{app:outflows}).}
\end{deluxetable}

Cold, $T\lesssim20$ K, and dense, \nHH~$\gtrsim10^4$ cm$^{-3}$, regions within molecular clouds provide the ideal conditions for stars and planets to form. Within dense clouds, the visual extinction, $A_{\rm V}$, becomes larger than about 3 mag so that the UV photon flux of the interstellar radiation field is attenuated.
There, cosmic rays (CRs) become primary ionising agents for the interstellar medium (ISM), also affecting the chemistry of the star-forming regions, regulating the coupling of the gas with the magnetic field (e.g. \citealt{Padovani20} and \citealt{Gabici22} as an overview).

Penetrating into molecular clouds, CRs ionize H$_2$ and produce several ions that are the starting point of ion-neutral chemistry. Among them, H$_3^+$ is directly formed by ${\rm H_2} + {\rm CR} \rightarrow {\rm H^+_2 + e^-}$, followed by the fast reaction ${\rm H_2^+} + {\rm H_2} \rightarrow {\rm H^+_3 + H}$. Due to its simple reaction chain, H$_3^+$ is the pivotal molecule to quantify the CR ionisation rate (CRIR) relative to molecular hydrogen (hereafter $\crir$) and to assess the role of CRs in star-forming environments (e.g., \citealt{Indriolo12}, \citealt{NeufeldWolfire17}; see also \citealt{Luo23a, Luo23b} for an alternative method recently proposed). Due to the lack of a permanent electric dipole, H$_3^+$, similar to H$_2$, does not emit rotational lines at cold temperatures. This method is hence limited to low $A_\mathrm{V}$, where H$_3^+$ can be observed in absorption towards bright infrared (IR) sources. Furthermore, since H$_3^+$ is widely spread in the ISM, disentangling the contribution of other clouds on the line-of-sight to the source can be challenging \citep[see e.g. the discussion in][]{VanDerTak00}.

Since \HHHp~is not observable, other tracers are needed to estimate $\crir$ in dense regions without strong background emission. Such methods were implemented in several works starting from the pioneering work of \cite{Black78} where a $\crir\sim10^{-17} \rm  s^{-1}$ was derived based on OH, CO, and HD. Analytical approaches, based on the steady-state assumptions and the abundances of DCO$^+$, HCO$^+$, and CO, have also been explored, reporting \mbox{$10^{-17}<\crir<10^{-14}\rm~s^{-1}$} \citep[e.g.][]{Caselli98}. These methods, however, depend on the number of independent tracers used to estimate the ionisation fraction, i.e. $x(e) = n(e)/n({\rm H_2})$, with $n(e)$ the free electron density. \cite{Ivlev19} derived a $\crir\sim10^{-16} \rm~s^{-1}$ with a pure theoretical method based on a self-consistent model in the prestellar core L1544. 

Over the years, multiple observational techniques have been proposed to estimate $\crir$ using different molecular tracers combined with chemical models predictions: HCO$^+$, N$_2$H$^+$ and their isotopologues (e.g. \citealt{Ceccarelli14b} and \citealt{Redaelli21b}), HC$_3$N and HC$_5$N (e.g., \citealt{Fontani17}) and c-C$_3$H$_2$ (e.g., \citealt{Favre18}). A very promising method based on H$_2$ near-infrared lines has been also recently proposed by \cite{Bialy20} and \cite{Padovani22}. Overall, there is no consensus on $\crir$ in dense regions. The various estimates differ by orders of magnitude, representing one of the most sensitive uncertainties in astrochemical models. 

In this Letter, we follow the new analytical approach proposed by \cite{Bovino20} 
to estimate $\crir$ in dense molecular clouds by using ortho-H$_2$D$^+$ (hereafter, \ohhdp) as the main observational constraint to derive the amount of \HHHp. In the following Sections, we report the details of this method and its limitations, then present our results and conclusions.

\section{Methodology}
The method proposed by \cite{Bovino20} is based on the following analytical formulation 
\begin{equation}\label{eq:crir}
    \crir = k^{{\rm H^+_3}}_{{\rm CO}}\:\frac{X({\rm CO})\:N(\text{o-}{\rm H_2D^+})}{3\:R_{\rm D}\:\ell}\,,
\end{equation} 
\noindent where $k^{{\rm H^+_3}}_{\rm CO}$ is the rate at which CO destroys H$^+_3$, $X$(CO) the abundance of CO relative to H$_2$, $R_\mathrm{D}$ is the deuteration of HCO$^+$ ($R_\mathrm{D}=$ DCO$^+$/HCO$^+$), $\ell$ is the path length over which the column densities ($N$) are estimated, and $N(\text{o-}{\rm H_2D^+})$ is the column density of the main H$_3^+$ isotopologue.

This methodology has shown to be accurate within a factor of 1.5-3 if the deuteration levels are well below 10\%, and the main H$_3^+$ isotopologue is H$_2$D$^+$. This means that once the latter is efficiently converted in D$_2$H$^+$ and D$_3^+$, the validity of the formula breaks. 
This method was applied to Atacama Pathfinder EXperiment (APEX) and IRAM-30m observations in a large sample of high-mass star-forming regions, yielding $\crir$ in the range of (0.7-6)$\times 10^{-17} \rm~s^{-1}$ \citep{Sabatini20}, and has the great advantage of being model independent.

In order to apply Equation~\ref{eq:crir} at core-scale, we then need to retrieve the column densities of the \HH, \ohhdp, \DCOp, HCO$^+$ and~CO from high-resolution observations.

\section{Source selection and data reduction}\label{sec2:sample}
\begin{figure*}
   \centering
   \includegraphics[width=1\hsize]{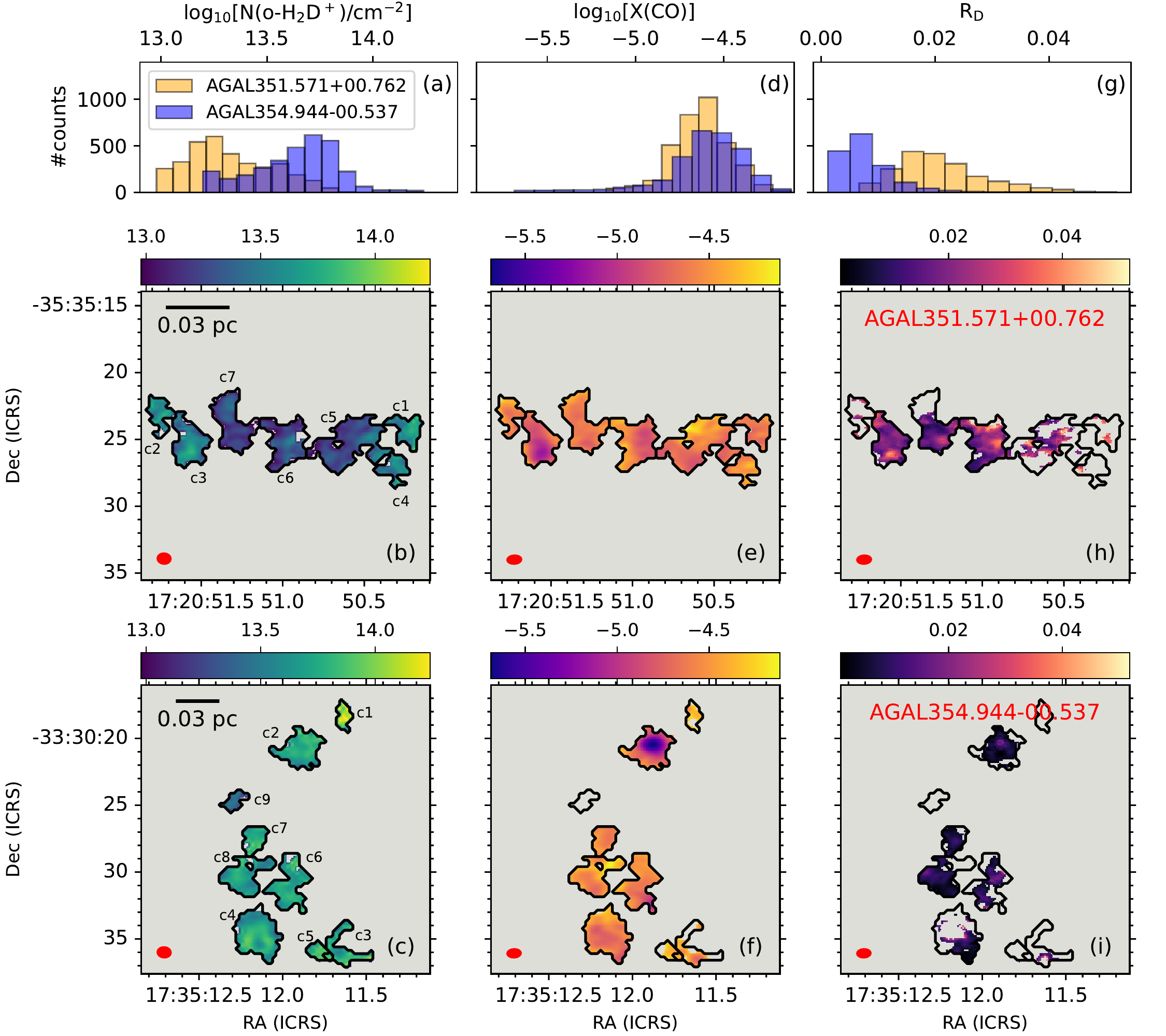}
\caption{(a) Orange histogram (reported as a map in panel b) shows the number distributions of log$_{10}$[$N$(\ohhdp)/cm$^{-2}$] derived for AG351. The blue histogram is the same for AG354 (mapped in panel $c$). Panels (d, e, f) are the same for log$_{10}$[X(CO)], while panels (g, h, i) refer to $R_{\rm D}$. Black contours in all the maps show the core's structures identified in \ohhdp by \cite{Redaelli21}. All the maps are masked with respect to the core structures. Cores' IDs (`c\#') follow the classification defined by \cite{Redaelli21} and are shown as black markers in panels (b, c). The ALMA synthesized beams are displayed in red in the lower left corner of each map, while the scale bars are shown in the upper left \mbox{corners of panels (b, c)}.}\label{fig:ingredients}%
\end{figure*}
The targeted sources are AGAL351.571+00.762 (hereafter AG351) and AGAL354.944-00.537 (AG354), which belong to the APEX Telescope Large Area Survey of the Galaxy (ATLASGAL; \citealt{Schuller09}) sample, which includes $\sim$10$^4$ massive clumps at various evolutionary stages. Their properties are reported in Table~\ref{tab:summary}. Both sources have been observed in \mbox{\ohhdp} with the Atacama Large Millimeter/submillimter Array (ALMA), at a resolution of $\approx$0.9\arcsec~($\sim$1400 AU; at the average distance of 1.6~kpc; \citealt{Redaelli21}). Those authors analysed the \ohhdp data via a dendrogram algorithm, identifying core-like structures that are believed to be truly prestellar due to the presence of \ohhdp emission (also reported in  Figure~\ref{fig:ingredients} $a, b,c$). This has also been confirmed by the absence of detected near-IR emission \citep{Kuhn21}. 

Additional ALMA observations were acquired during Cycle-8 (2021.1.00379.S; PI:~G.~Sabatini), using  the 12m-Array (Main Array) and the Atacama Compact Array (ACA), including both the 7m-Array and the Total Power (TP). The 12m-array included 43-47 antennas with baselines between 15 and 500~m, whilst ACA 9-10 antennas,  distributed over baselines range of 9-45~m. All the observations were carried out with a precipitable water vapour $< 2.5$~mm.

The observations made use of the single-point mapping mode and two separate spectral setups (SeS) with a spectral resolution $\Delta\nu=0.1$ km s$^{-1}$. The first SeS covers the \DCOp (3-2) and \CeigO (2-1) lines at $\sim 216.11\, \rm GHz$ and $\sim219.56\, \rm GHz$, respectively. The average angular scales covered at these frequency ranges from a resolution $\sim$0.7\arcsec$\times$1\arcsec~to a maximum recoverable scale of $\sim$32\arcsec, i.e. $\sim$(1-50)$\times$10$^3$~AU at the average distance of 1.6~kpc. The second SeS covers the \HtrCOp (3-2) line at $\sim260.25\, \rm GHz$ (angular scales from $\sim$0.5\arcsec~to $\sim$26\arcsec).

All data were calibrated with the pipeline in CASA~6.2.1-7, while CASA-6.4 was employed for imaging, using the \textsc{tclean} task. The continuum images were obtained by averaging the line-free channels in an effective bandwidth of $\sim$1.4$\,$GHz around 224$\,$GHz. The final beam size of the continuum maps was $\sim$0.7\arcsec$\times$1\arcsec. Line cubes were made assuming a briggs-robust = 0.5 weighting, and a multiscale option (scales: 0, 5, 15, and 50 times the pixel size). All cubes were generated with the `auto-multithresh' algorithm, with the exception of C$^{18}$O, for which we used a manual masking procedure, due to the extension of its emission. The images have 384$\times$384 pixels with a pixel size of 0.2\arcsec. Finally, the 12 and 7~m array line cubes were combined with the TP observations through the classic feathering technique. The typical 1$\sigma$ rms are summarised in Table~\ref{tab:summary}, whilst additional details are reported in Appendix~\ref{app:outflows}.

To correct for the absence of zero-scales in the continuum maps and \ohhdp cubes, we used the ratio of the total flux recovered with ALMA and the one observed with the single-dish APEX, using the ATLASGAL data for the continuum and the APEX spectra in \cite{Sabatini20} for \ohhdp.

\section{Analysis and results}\label{sec3:analisys}
Since the main CO and HCO$^+$~isotopologues are almost always optically thick \citep[e.g.,][]{Heyer15}, their intensity is not proportional to $N$. Therefore, in our analysis, we have used less abundant isotopologues (i.e., \CeigO~and \HtrCOp) to obtain a much more accurate estimate of $N$. In this Section, we summarise the procedure and the assumptions we follow to derive $\crir$.

\subsection{{\rm \HH}~column density maps}\label{sec3.1:NHH}
The maps of the beam-averaged \HH~column density are computed following \cite{Sabatini22}, using the dust temperature values listed in Table~\ref{tab:summary}. We used a standard propagation of the uncertainties on the ALMA continuum flux, finding an average uncertainty of $8.5\times10^{21} \rm \; cm^{-2}$.
The average $N({\rm H}_2)$ values for each core span the range (1-2.5)$\times~10^{23} \rm~cm^{-2}$, and are consistent with previous measurements at 0.8 mm \citep{Redaelli21}.

%


\subsection{Column density maps of the molecular tracers}\label{sec3.2:Ntracers}
The column densities of all the molecular species are estimated following \cite{Mangum15}, computing optical depths, $\tau$, with Equation~5 of \cite{Caselli08}. All the molecular parameters are taken from the Cologne Database for Molecular Spectroscopy\footnote{CDMS: \url{https://cdms.astro.uni-koeln.de/classic/}.}, while the partition functions at the excitation temperature ($T_\mathrm{ex}$) were obtained by linearly interpolating the values reported by \cite{Sabatini20} for \ohhdp, \cite{Redaelli19} for \DCOp, and the CDMS catalogue for \CeigO~and \HtrCOp. The integral of the optical depth along the velocity axis is computed in each channel where the emission is $>3\sigma$ (see Table~\ref{tab:summary}).
\begin{figure*}
   \centering
   \includegraphics[width=1\hsize]{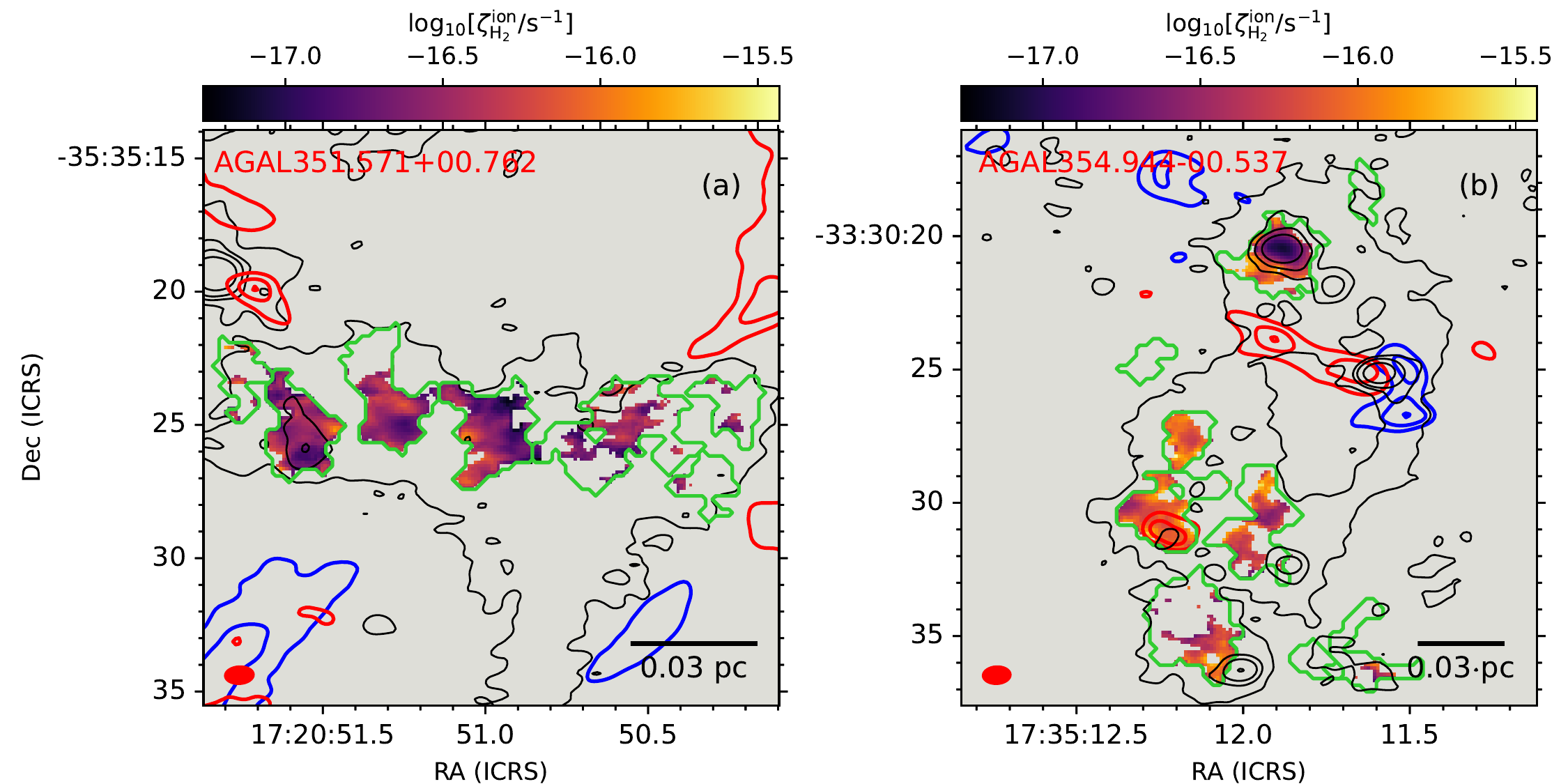}
\caption{Map of the log$_{10}(\crir)$ derived for AG351 (panel $a$) and AG354 ($b$). The black contours show the continuum emission at $1.33\, \rm mm$ (levels: $[3, 6, 9, 15, 30]\sigma$, whit 1$\sigma = 0.1 \rm~mJy\,beam^{-1}$). The blue- and red-shifted components of $^{12}$C$^{16}$O emission are shown as blue and red contours, respectively, whit levels at 50\%, 75\%, and 99\% of the peak of the CO velocity-integrated intensity. More details in Appendix~\ref{app:outflows}. Green contours mark the same core structures reported in Figure~\ref{fig:ingredients}. The $\crir$ was derived only in those positions where all the molecular tracers have been detected at $>3\sigma$ (see Tab.~\ref{tab:summary}). The beam size and the scale bar are shown in the bottom-left and bottom-right corners, respectively.}\label{fig:crirmaps}%
\end{figure*}

In order to select the $T_\mathrm{ex}$ value for each tracer, we analyse their different chemical properties (i.e. different critical densities, and spatial distributions). For the \ohhdp~we used $T_\mathrm{ex}=10$~K as found by \cite{Redaelli21}. We assumed the same excitation temperature also for \DCOp and \HtrCOp~considering their chemical connection with \ohhdp~(e.g. \citealt{Dalgarno84}). We note, however, that the estimate of $\crir$, obtained via Equation~\ref{eq:crir}, is not directly affected by the value of $N$(\DCOp) and $N$(\HtrCOp), but rather by their ratio. Since the two species have similar molecular properties and the same transition applies to both, their abundance ratio is not strongly dependent on $T_{\rm ex}$. A variation of 5 to 20~K in $T_{\rm ex}$ would imply a difference of $\pm20\%$ in $N$(\DCOp)/$N$(\HtrCOp), which is consistent with the final average error associated with the $\crir$ (see Section~\ref{sec3.3:crir}). The \CeigO~has relatively lower critical densities ($n_c \sim 10^3 \rm~cm^{-3}$). Since we are focusing on high-density regions, it is reasonable to assume that \CeigO~is thermalised by collisions with \HH, and that gas and dust are coupled \citep{Goldsmith01}. We hence assume $T_\mathrm{ex} \mathrm{(C^{18}O)}= T_\mathrm{kin} = T_\mathrm{dust}$. At these temperatures, the emission of all the tracers is mostly optically thin ($\tau<1.1$), and the final $N$ are consistent with those typically obtained in similar sources (e.g. \citealt{Roberts11, Morii21, Sabatini22}). 
\subsection{Core-scale $\crir$ maps}\label{sec3.3:crir}
We derived $X$(CO) from the \CeigO~abundance assuming an oxygen isotopic ratio ${\rm [^{16}O]/[^{18}O]} = 58.8R_{\rm GC} + 37.1$, where $R_{\rm GC}$ is the galactocentric distance of each source expressed in kpc \citep{WilsonRood94}. 
The resulting $X$(CO) span more than one order of magnitude, from $\sim7\times10^{-6}$ to $6\times10^{-5}$ within the individual cores (see Figures~\ref{fig:ingredients} $d,e,f$). We identify regions where the CO-depletion (\fD) is almost irrelevant, with observed abundances of \CeigO~as expected (see \citealt{Giannetti17_june}), up to regions where only less than 5\% of the expected CO is still present in the gas phase {\bf (i.e., \fD~$>20$)}. These values agree with the average \fD~found by \cite{Sabatini22} in a large sample of prestellar cores candidates embedded in young high-mass star-forming regions. For $R_{\rm D}$, we derived $N$(HCO$^+$) from \HtrCOp~assuming the ratio ${\rm [^{12}C]/[^{13}C]} = 6.1R_{\rm GC} + 14.3$ \citep[e.g.][]{Feng16}. The resulting $R_{\rm D}$ are in between 0.002 and 0.05 (Figures~\ref{fig:ingredients} $g,h,i$), slightly higher than those previously found by \cite{Sabatini20} with APEX single-dish observations, and most likely due to beam-dilution effects.

In Equation~\ref{eq:crir}, we employ $k^{{\rm H^+_3}}_{\rm CO}\sim2.3\times10^{-9}\, \rm cm^{3} \, s^{-1}$, which is derived at the $T_{\rm gas}=T_{\rm dust}$ ($\sim18$~K)\footnote{The rate is taken from the KInetic Database for Astrochemistry (KIDA); \url{https://kida.astrochem-tools.org/} with modifications based on \citet{Sipila15} to take into account different isomers.}. We highlight that a significant temperature variation (down to e.g. $T_\mathrm{gas} =10\, \rm K$) leads to an increase of the reaction rate of less than $9$\%. As we account for the large-scale emission of all the tracers by including ALMA-TP in Equation~\ref{eq:crir} we have also assumed $\ell$ equal to the ALMA Field of View.

The final $\crir$ maps are shown in Figure~\ref{fig:crirmaps} together with the thermal continuum emission at 1.3~mm. The $\crir$ spans from $\sim 6\times10^{-18}$ to $2\times10^{-16}\rm~s^{-1}$, and shows a global pattern within the cores identified in \ohhdp. In particular, we note that where the continuum emission is higher, the average $\crir$ tends to decrease (see Section~\ref{sec4:discussion_conclusions}). The $\crir$ estimates increase by a factor of 2 at most if we restrict our analysis to the scales covered by the interferometric data, i.e. removing the TP contribution from all tracers and taking the average core size in each source (see Appendix~\ref{app:outflows}) as the representative value for $\ell$ in Equation~\ref{eq:crir}. Assuming a standard propagation of the uncertainties on the main beam temperature of each tracer observed with ALMA, the typical uncertainties on the $\crir$ estimates are in between a factor of 1.1 and 1.7.

\section{Discussion and Conclusions}\label{sec4:discussion_conclusions}
 \begin{figure}
   \centering
   \includegraphics[width=1\hsize]{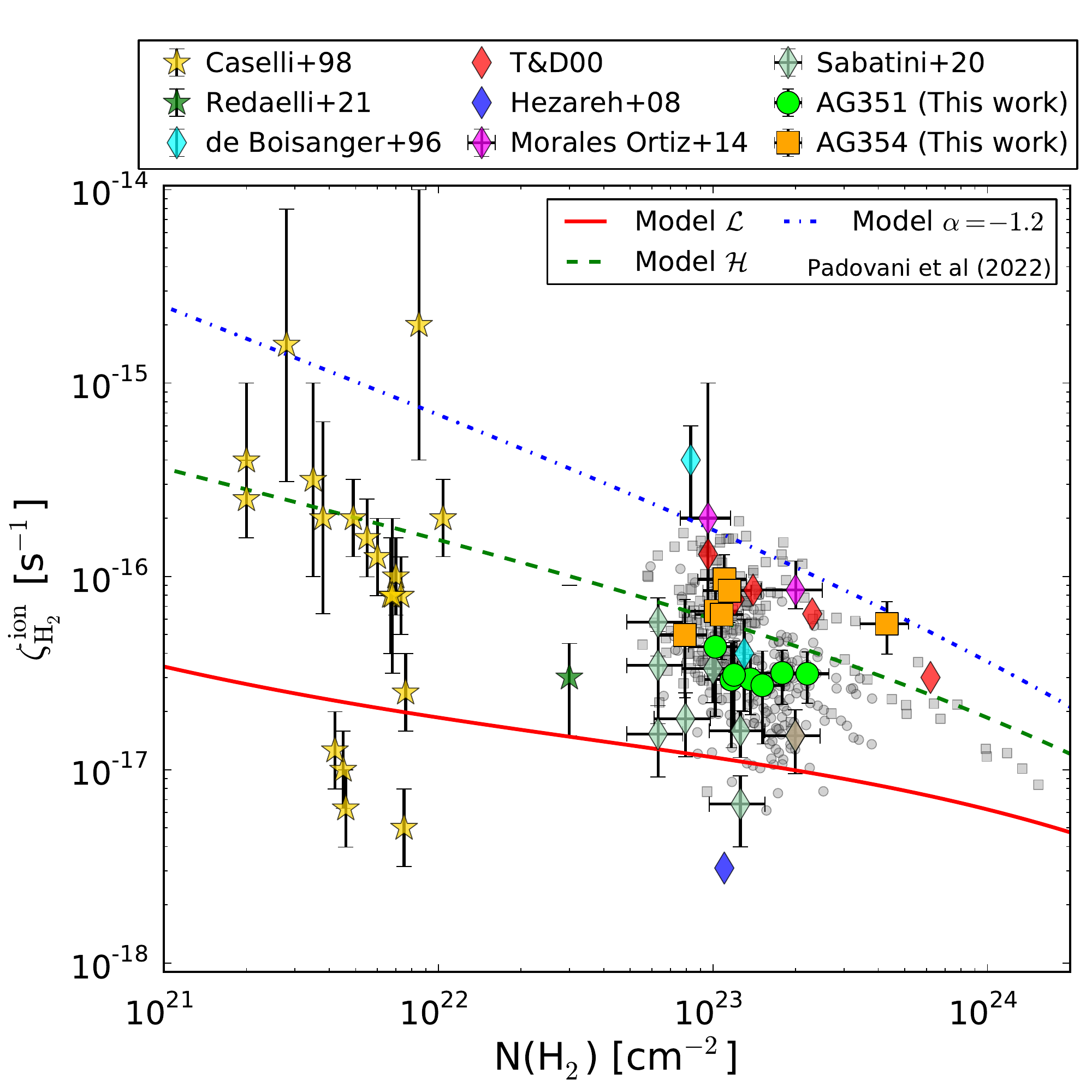}
\caption{Estimated $\crir$ vs $N$(H$_2$) for a sample of low-mass (stars; $yellow$ from \citealt{Caselli98}, $green$ from \citealt{Redaelli21b}) and high-mass star-forming region (diamonds; $cyan$ from \citealt{deBoisanger96}; $red$ from \citealt{VanDerTak00} (T\&D00; see also \citealt{Sabatini20}), $blue$ from \citealt{Hezareh08} and $magenta$ from \citealt{MoralesOrtiz14}). Grey circles and squares refer to our estimates of $\crir$ in each resolution element in Figure~\ref{fig:crirmaps}, for AG351 and AG354 respectively, whilst green circles and orange squares are the average values for each core identified in \ohhdp. The relative uncertainties on the grey circles and squares are in between 10\% and 70\% (Section~\ref{sec3.3:crir}). Dashed/dotted lines show the models discussed in \cite{Padovani22} considering different slopes for the CR proton spectrum (see discussion in Section~\ref{sec4:discussion_conclusions}).}\label{fig:crirVSNH2}%
\end{figure}
To the best of our knowledge, Figure~\ref{fig:crirmaps} provides the first look at the $\crir$ distribution in two intermediate-mass star-forming regions observed at the remarkable angular resolution of ALMA. AG351 and AG354 are two relatively massive clumps that are similar in terms of physical properties (i.e. total clump's mass, $\langle\Tdust\rangle$~and $R_{\rm GC}$), but far enough apart to be considered independent (Table~\ref{tab:summary}). This gives us a double advantage. On the one hand, by looking at the maps in Figure~\ref{fig:crirmaps} individually, we can assess the local variation of the $\crir$ in two separate samples of low-mass prestellar cores harboured in different parental environments \citep{Redaelli21}. With this approach, we reduce the impact of the initial chemical and physical conditions of the gas that originally formed the two clumps on the $\crir$. On the other hand, with two $\crir$ maps, we can also compare the $\langle\crir\rangle$ in different cores' samples, assessing the impact of the above initial conditions on the evolution, and the distribution, of $\crir$. 

To better discuss the above points, we report in Figure~\ref{fig:crirVSNH2} the estimates of $\crir$ obtained for a sample of high- (diamonds) and low-mass star-forming regions (stars)\footnote{A similar plot is also provided in Appendix~\ref{app:comparison}, where we made a comparison with other analytical methods.}. The new $\crir$ estimates obtained with ALMA are shown as grey circles (AG351) or squares (AG354). Each grey dot corresponds to a single resolution element in Figure~\ref{fig:crirmaps} and thus has statistical significance, with typical uncertainties in between 10\% and 70\% (see Sect.~\ref{sec3.3:crir}). Following \cite{Sabatini22}, we have assumed a conservative error of 20\% in $N$(\HH), which corresponds to the typical uncertainty when $N$(\HH) is derived from ALMA continuum observations.
Green circles and orange squares are the $\langle\crir\rangle$ derived for the individual cores in AG351 and AG354, respectively. A comparison of the $\langle\crir\rangle$ obtained in the two sources shows that the cores belonging to the same parental clump display comparable values within the error bars. This suggests that the ionisation properties are set by the global properties of the environment where the cores formed.

Our results also show a possible trend when comparing the two core populations, in that $\crir$ in AG354 appears a factor of $\approx 2$ higher than in AG351. This might hint at distinct initial conditions for the two clumps, or at slightly different evolutionary scenarios. Almost all prestellar cores identified in the two clumps via \ohhdp~have a virial parameter $< 1$ and, unless supported by strong magnetic fields, appear to be gravitationally unstable against collapse (\citealt{Kauffmann13}). As pointed out by \citet[][]{Redaelli21}, AG354 also reaches higher $N$(\HH) values and degree of turbulence, with larger linewidths. As a result of the progressive accretion of material from clumps to cores, the core population in AG354 is associated with larger $M_{\rm core}$ and $N$(\ohhdp), while having a comparable $X$(\ohhdp) as in AG351 (see Table~\ref{tab:cores_properties}).

We have also investigated the presence of outflows in the clumps using the CO (2-1) emission (see Appendix~\ref{app:outflows}). We find no significant high-velocity emission in AG351, whilst we detect a possible bipolar outflow in AG354. This seems associated with a continuum core lacking \ohhdp, and it is directed in the east-west direction, without overlapping with any \ohhdp-identified structure. We can therefore speculate that the different $\langle\crir\rangle$ derived in AG351 and AG354 is most likely due to the different amounts of material interacting with CRs. With the available information, however, we are not able to disentangle whether the different $\crir$ that we derived in the two clumps have caused a distinct evolutionary speed, or if they are on the contrary, influenced by e.g. the presence of protostellar activity. Furthermore, it is important to highlight that the detected difference is within the accuracy of the method itself.

Figure~\ref{fig:crirVSNH2} also shows the trends predicted by CR propagation models (see \citealt{Padovani18, Padovani22}). We report three reference cases, obtained considering a single CR electron spectrum and different CR proton spectra: ($i$) Model \modelL~ is based on the data from the two Voyager spacecraft (e.g.~\citealt{Stone19}) and considers a low-energy spectral slope $\alpha = 0.1$; ($ii$) Model \modelH, $\alpha = -0.8$, reproduces the average value of $\crir$ in diffuse regions (see also \citealt{Padovani20, Gabici22}); ($iii$) the model with $\alpha = -1.2$, which can be considered as an upper limit for the $\crir$ estimates in diffuse regions \citep{Padovani22}. Comparing the predictions of the model with our estimates, we found a global variation of $\crir$ with $N$(\HH), which is predicted by the CR propagation models and confirmed by observations. In both cases, the higher the $N$(\HH) values, the lower and more concentrated the $\crir$ estimates. Our results are in remarkable agreement with the latest models of CR propagation, whilst the scatter of $\crir$ values derived at a given $N$(\HH)~could reflect a different morphology of magnetic fields. This effect is indeed expected from theoretical models \citep[e.g., ][]{Padovani11,Padovani13}. The increasing capabilities of modern astronomical facilities, e.g. ALMA, in polarisation observations will allow us in the near future to study the magnetic field properties of the targeted sources to reconstruct the complete picture of the ionisation and dynamical properties of star-forming regions.

\begin{acknowledgments}
     The authors thank the anonymous referee for her/his suggestions to improve the manuscript, and gratefully acknowledge Dr M.~Padovani and Dr D.~Petry for fruitful discussions and feedback. SB is financially supported by ANID Fondecyt Regular (project \#1220033), and the ANID BASAL projects ACE210002 and FB210003. ER acknowledges the support of the Max Planck Society. The publication has  received funding from the Italian node of the European ALMA Regional Centre and also from the European Union's Horizon 2020 research and innovation program under grant agreement No 101004719 (ORP). This Letter makes use of the following ALMA data: ADS/JAO.ALMA\#2021.1.00379.S (PI: G.~Sabatini). ALMA is a partnership of ESO (representing its member states), NSF (USA) and NINS (Japan), together with NRC (Canada), MOST and ASIAA (Taiwan), and KASI (Republic of Korea), in cooperation with the Republic of Chile. The Joint ALMA Observatory is operated by ESO, AUI/NRAO and NAOJ.  
\end{acknowledgments}

\facilities{Atacama Large Millimeter/submillimeter Array; Atacama Pathfinder EXperiment.}

\software{This research has made use of \href{https://aplpy.readthedocs.io/en/stable/}{{\verb~APLpy~}} (an open-source plotting package for Python), \href{http://www.astropy.org}{\verb~Astropy~}, \href{https://numpy.org/}{\verb~NumPy~}, \href{https://matplotlib.org/}{\verb~Matplotlib~}, the Cologne Database for Molecular Spectroscopy (\href{https://cdms.astro.uni-koeln.de/}{CDMS}),  the KInetic Database for Astrochemistry (\href{https://kida.astrochem-tools.org/}{KIDA}), and the NASA’s Astrophysics Data System Bibliographic Services (\href{https://ui.adsabs.harvard.edu/}{ADS});}

\begin{deluxetable*}{l|cccccccccc}
\savetablenum{2}
\tablewidth{\columnwidth}
\caption{Physical properties of the cores harbored in AG351 and AG354\label{tab:cores_properties}}
\tablehead{
Core-ID\tablenotemark{\small{a}} & $R_{\rm eff}$\tablenotemark{\small{a}} & $M_{\rm vir}$\tablenotemark{\small{a}} & $M_{\rm core}$\tablenotemark{\small{a}} & $N$(\HH) &  $N$(\ohhdp) & $R_{\rm D}$ & $R_{\rm H}$ & X(CO) & \fD & $\crir$\\
  & (AU) & (M$_\odot$) & (M$_\odot$) & ($\times10^{23}$~cm$^{-2}$) & ($\times10^{13}$~cm$^{-2}$) & ($\times10^{-2}$) & ($\times10^{-5}$) & ($\times10^{-5}$) & & ($\times$10$^{-17}\rm~s^{-1}$)}
 \startdata
(AG351) c1 & 1500 & $0.4\pm0.1$ & $0.7\pm0.2$ & $1.5\pm0.3$ & $3.8\pm0.9$ & $3.4\pm1.6$ & $0.3\pm 0.1$ & 2.3 & 5.4 & $2.7\pm 1.4$\\
(AG351) c2 & 1400 & $0.7\pm0.3$ & $0.5\pm0.2$ & $1.0\pm0.2$ & $3.3\pm0.8$ & $2.7\pm1.4$ & $0.4\pm 0.1$ & 3.0 & 4.4 & $4.3\pm 2.4$\\
(AG351) c3 & 2700 & $1.0\pm0.4$ & $1.8\pm0.6$ & $2.2\pm0.4$ & $3.2\pm0.6$ & $2.1\pm0.6$ & $0.7\pm 0.1$ & 1.6 & 8.8 & $3.1\pm 0.9$\\
(AG351) c4 & 1400 & $0.4\pm0.2$ & $0.4\pm0.1$ & $1.2\pm0.2$ & $3.0\pm0.9$ & $3.4\pm1.8$ & $0.3\pm 0.1$ & 2.8 & 4.3 & $2.9\pm 1.6$\\
(AG351) c5 & 2700 & $0.6\pm0.1$ & $1.6\pm0.5$ & $1.2\pm0.2$ & $1.9\pm0.2$ & $2.3\pm1.1$ & $0.4\pm 0.1$ & 2.8 & 4.8 & $3.1\pm 1.5$\\
(AG351) c6 & 2100 & $1.1\pm0.3$ & $1.1\pm0.4$ & $1.4\pm0.3$ & $1.8\pm0.2$ & $2.2\pm0.7$ & $0.6\pm 0.1$ & 2.4 & 5.8 & $2.9\pm 1.0$\\
(AG351) c7 & 2200 & $0.5\pm0.1$ & $1.6\pm0.5$ & $1.8\pm0.4$ & $1.7\pm0.2$ & $1.6\pm0.4$ & $0.6\pm 0.1$ & 2.3 & 5.3 & $3.2\pm 1.0$\\
\tableline
(AG354) c2 & 2900 & $0.8\pm0.2$ & $5.6\pm1.8$ & $4.3\pm0.9$ & $5.0\pm 0.8$ & $0.7\pm0.1$ & $3.3\pm0.5$ & 1.1 & 16.8 & $5.7\pm1.7$ \\
(AG354) c3 & 2300 & $0.7\pm0.3$ & $1.1\pm0.4$ & $0.8\pm0.2$ & $5.1\pm 1.2$ & $2.1\pm1.0$ & $0.7\pm0.2$ & 2.9 & 4.3  & $5.0\pm2.6$ \\
(AG354) c4 & 3300 & $1.6\pm0.5$ & $3.0\pm1.0$ & $1.1\pm0.2$ & $4.9\pm 0.7$ & $1.3\pm0.6$ & $1.3\pm0.4$ & 2.3 & 5.3  & $6.3\pm2.6$ \\
(AG354) c6 & 2900 & $2.0\pm0.4$ & $2.2\pm0.7$ & $1.0\pm0.2$ & $4.8\pm 0.4$ & $1.5\pm0.5$ & $0.8\pm0.2$ & 2.6 & 4.6  & $6.6\pm2.6$ \\
(AG354) c7 & 1900 & $1.5\pm1.3$ & $1.2\pm1.2$ & $1.1\pm0.2$ & $5.4\pm 0.6$ & $1.0\pm0.3$ & $0.9\pm0.3$ & 2.8 & 4.2  & $9.6\pm3.3$ \\
(AG354) c8 & 2800 & $1.1\pm0.2$ & $2.1\pm0.7$ & $1.2\pm0.2$ & $4.2\pm 0.3$ & $1.1\pm0.3$ & $1.0\pm0.2$ & 3.1 & 3.9  & $8.4\pm2.7$ \\
 \enddata
\tablecomments{$^{(a)}$ IDs and values taken from \cite{Redaelli21}; see also Figure~\ref{fig:ingredients}. For the sake of clarity, the Table only summarises the average properties of the cores for which $\crir$ is available. From left to right: effective radius; virial mass; total gas mass; average column density of \HH; average \ohhdp~column density; deuterium fraction, $N$(DCO$^+$)/$N$(HCO$^+$); hydrogenation fraction, $N$(HCO$^+$)/$N$(CO); CO/\HH~abundance from \CeigO, with a typical uncertainty of $\lesssim$10\% derived considering the average error in the $R_{\rm GC}$ of each source (see \citealt{Urquhart18}); CO-depletion factor (typical uncertainty of $\sim$15\%; see \citealt{Sabatini22}); average $\crir$ as derived in Section~\ref{sec3.3:crir}.}
\end{deluxetable*}

\appendix
\section{Core properties and molecular outflows identification}\label{app:outflows}

This Appendix summarises the physical properties derived for the core populations that are harboured in AG351 and AG354. In Table~\ref{tab:cores_properties} we report the average values of several physical and chemical quantities of the cores. For some of them -- e.g. the effective radius ($R_{\rm eff}$), and the virial and total gas masses ($M_{\rm vir}$ and $M_{\rm core}$, respectively) -- we refer to \cite{Redaelli21} for a comprehensive description of their derivation. All the other quantities were obtained by averaging the value shown, or used to derive the parameters, in Figures~\ref{fig:ingredients}~and~\ref{fig:crirmaps}.

As an additional view on the dynamical state of the gas globally involved in the clumps, we have investigated the presence of protostellar molecular outflows using the CO (2-1) emission. The CO data are part of the ALMA campaign described in Section~\ref{sec2:sample}. We observed the $^{12}$C$^{16}$O (2-1) line at~$\sim$230.54~GHz in a third SeS centred at 230.5~GHz. The data were taken with a spectral resolution of 0.6 km~s$^{-1}$, which is six times larger than that of the other tracers, to maximise the sensitivity and provide a good detection of the thermal continuum emission. The calibration and the imaging were performed in accordance with Section~\ref{sec2:sample}. Figure~\ref{fig:crirmaps} shows the blue- and red-shifted components (blue and red contours, respectively) of the CO emission (12m+7m+TP), superimposed on the $\crir$ maps. The blue and red-shifted components were derived by integrating the CO emission in the velocity ranges between $\sim$[-20, -6] km s$^{-1}$ and $\sim$[-2, 35] km s$^{-1}$, respectively. The presence of protostellar molecular outflows seems evident only in core-2 of AG354 (see Figure~\ref{fig:crirmaps} and Table~\ref{tab:cores_properties}). In this case, the outflow has a projected size of a few arcseconds, but it never overlaps with any of the structures where the column densities and $\crir$ were derived. On the contrary, AG351 lacks any evidence of protostellar activity.

\section{Comparison with previous methods}\label{app:comparison}
\begin{figure*}
   \centering
   \includegraphics[width=0.9\hsize]{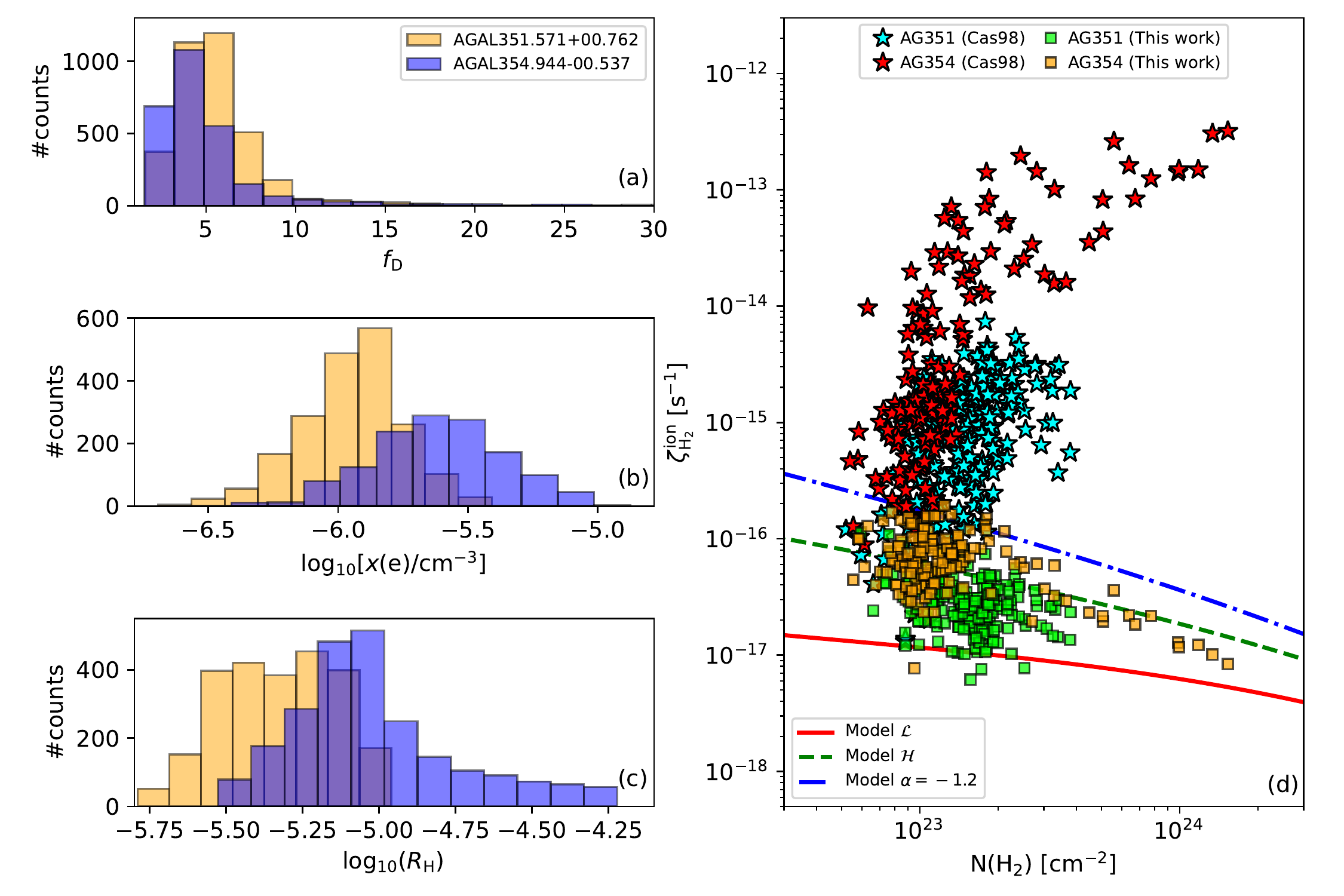}
\caption{Panels (a,b,c) show the number distributions of \fD, $x$($e$) and $R_{\rm H}$, respectively (orange for AG351 and blue for AG354). Panel (d) compares the $\crir$ derived using the method of \cite{Caselli98} (stars) or  \cite{Bovino20} (squares) as a function of $N$(\HH). Colours refer to different sources, while the line profiles are the same models discussed in Figure~\ref{fig:crirVSNH2}.
}\label{fig:comparison}%
\end{figure*}

In this Appendix, we provide results obtained with the analytical method proposed by \cite{Caselli98}. 
We note that already in \citet{Caselli98}, the authors avoided the use of this method and preferred to run chemical models to estimate the $\crir$ in several low-mass cores. The limitations of this formulation are also discussed in \citet{Caselli02} and lie primarily in the lack of some terms in the chemistry of H$_2$D$^+$ and the reactions involving electrons, and it appears to greatly overestimate $x$($e$) and the associated $\crir$ . However, recent studies have used this method to estimate the $\crir$ in a protostellar source \citep{Cabedo23}, where the physical conditions are far from the range of applicability of the original formula (determined at 10 K and for early stages of star-forming regions). 
For the Bok Globule B335, which hosts a Class 0 protostar with already developed outflow, \citet{Cabedo23} obtained a $\crir\sim$7$\times$$10^{-14}$ s$^{-1}$, by assuming a $T_{\rm ex}=25$~K for each species. 
They explain the resulting $\crir$ as a local effect of the protostellar activity, which however, would also affect \fD, that will decrease at $T_{\rm gas}>20$~K, when CO is efficiently desorbed in the gas phase. However, their resulting \fD~present values of about 80 at the position of the protostellar embryo, i.e. an unexpected extremely high depleted region. It is also worth noting that any other process that might affect the estimates of \fD~(such as the UV photodissociation of CO due to protostellar activity or the CO conversion to other species such as HCO$^+$ by CRs) would prevent the application of the method proposed by \cite{Caselli98}, since the underlying assumptions imply that only a fraction of C and O (i.e., 1/\fD) remains in the gas phase, while the rest is frozen, in the form of CO, on the surface of the grains.
To understand the results reported by \citet{Cabedo23}, we then decided to apply the  formula proposed by \citet{Caselli98} to our sources, which fall in the correct range of applicability being prestellar in nature. From the comparison between a prestellar and a protostellar source we can then inspect the validity of the method and its reliability.

Figure~\ref{fig:comparison} summarises the results of this test: Panel ($a$) shows the \fD~distributions derived from \CeigO, assuming a canonical CO abundance which varies with $R_{\rm GC}$ and the ${\rm [^{16}O]/[^{18}O]}$ ratio in Sect.~\ref{sec3.3:crir}. The CO-depletion factors derived in this way always verify the condition of applicability for \cite{Caselli98}, which requires $R_{\rm D} < 0.023\times f_{\rm D}$; Panel ($b$) shows the $x$($e$) obtained from Equation~3 in \cite{Caselli98}; Panel ($c$) shows the hydrogenation fraction ($R_{\rm H}$) from the HCO$^+$/CO ratio. We stress that the ranges of \fD~and $R_{\rm H}$ reported in Figure~\ref{fig:comparison} are consistent with what is typically reported in the literature for similar sources (e.g. \citealt{Caselli98, Sabatini22}), while we find $x$($e$) a factor of 10-100 larger (see \citealt{deBoisanger96, Caselli02b}). In Figure~\ref{fig:comparison}~($d$) we show the comparison between the $\crir$ obtained following \cite{Bovino20} (squares) and \cite{Caselli98} (stars). The latter span values in between 10$^{-16}$-10$^{-13}$ s$^{-1}$, i.e. from one to two orders of magnitude larger than the ones obtained with the method of \citet{Bovino20}. In addition, they tend to overestimate the upper limits provided by the CR propagation model with $\alpha=-1.2$. This discrepancy also increases when moving to larger $N$(\HH). 

All the values reported in Figure~\ref{fig:comparison} ($a,b,c$) are similar to those obtained by \citet{Cabedo23}, even if the analysed samples show very different physical conditions. From this result, 
we conclude that the method proposed by \cite{Caselli98} shows significant limitations in its applicability. For a more in-depth discussion of the limits of the methods, we refer to \cite{Caselli02}.

\clearpage

\bibliography{mybib_GAL}{}
\bibliographystyle{aasjournal}

\end{document}